\documentclass[a4paper,11pt]{article}
\usepackage{pos}

\newcommand{\ud}{\mathrm{d}}

\title{Electromagnetic and gravitational form factors of the nucleon}
\author*[a]{C\'edric Lorc\'e}

\affiliation[a]{CPHT, CNRS, \'Ecole polytechnique, Institut Polytechnique de Paris, \\
91120 Palaiseau, France}

\emailAdd{cedric.lorce@polytechnique.edu}

\abstract{Form factors are Lorentz invariant functions describing the internal structure of a system. In particular, they encode how physical properties like, e.g., charge, energy, momentum, and pressure are spatially distributed. While nucleon electromagnetic form factors have been studied for a long time, the first extraction of nucleon gravitational form factors from experimental data was reported in 2018, triggering a lot of enthusiasm and attention in the hadronic community. In this contribution we review some theoretical bases, discuss recent developments regarding the physical interpretation of these form factors, and give a glimpse of what can be learned about the mass and spin structure of the nucleon.}

\FullConference{25th International Spin Physics Symposium (SPIN 2023)\\
 24-29 September 2023\\
 Durham, NC, USA\\}

\begin{document}
\maketitle

\section{Introduction}

Exclusive scattering experiments allow one to probe how physical properties of a microscopic system are distributed in space. In particular, the electromagnetic structure of hadrons has been measured with extremely high precision over the past decades, see e.g.~\cite{Punjabi:2015bba,Gao:2021sml}~for recent reviews. Interestingly, it turns out that the mechanical structure of hadrons can also be measured, albeit indirectly~\cite{Burkert:2023wzr}.

The study of elastic lepton-hadron scatterings provides information on the matrix elements of the electromagnetic current $\langle p',s'|\hat j^\mu|p,s\rangle$, where $p$ is the four-momentum and $s$ is the polarization of the hadron. These matrix elements can be parametrized in terms of Lorentz-invariant functions known as electromagnetic form factors (FFs), and are usually interpreted after Fourier transform in the Breit frame (BF) in terms of charge and magnetization densities~\cite{Ernst:1960zza,Sachs:1962zzc}. However, because of relativistic recoil corrections, these cannot be considered as probabilistic distributions~\cite{Yennie:1957skg,Burkardt:2000za,Jaffe:2020ebz}. Genuine charge and magnetization densities can alternatively be defined on the light front (LF)~\cite{Burkardt:2002hr,Miller:2010nz,Miller:2018ybm}, but they display distortions that are hard to conciliate with the picture of a system at rest~\cite{Miller:2007uy,Carlson:2007xd}.

Various attempts to clarify the concept of relativistic spatial distribution have recently appeared in the literature, see e.g.~\cite{Panteleeva:2021iip,Kim:2021kum,Freese:2021mzg,Epelbaum:2022fjc,Panteleeva:2022khw,Li:2022hyf,Freese:2022fat,Freese:2023abr}, but the only way that reconciles BF and LF distributions while explaining at the same time the origin of LF distortions is to adopt a phase-space approach~\cite{Lorce:2018zpf,Lorce:2018egm,Lorce:2020onh}. The latter allows one to embrace the frame dependence of relativistic spatial distributions by relaxing the density interpretation to a quasi-probabilistic one. Some of the results obtained within this approach are summarized in the rest of this contribution.

%%%%%%%%%%%%%%%%%%%%%%%%%%%%%%%%%%%%%%%%%%%%%%%%%%%%%%%%%%%%%%%%%%%%%%%%%%%%%%%%%%
\section{Phase-space formalism}

The physical meaning of LF distributions can usually be better understood in terms of the closely related distributions in the infinite-momentum frame (IMF)~\cite{Kogut:1969xa}, where the target moves with almost the speed of light relative to the inertial observer. To interpolate between the BF and IMF descriptions, the concept of elastic frame (EF) distribution was first introduced in~\cite{Lorce:2017wkb} and further motivated by the phase-space approach~\cite{Lorce:2018zpf,Lorce:2018egm}. 

The expectation value of a generic operator $O$ in a physical state $|\psi\rangle$ can be represented in phase-space as follows
\begin{equation}
    \langle \psi|O|\psi\rangle=\int\frac{\ud^3P}{(2\pi)^3}\,\ud^3R\,\rho_\psi(\vec R,\vec P)\,\langle O\rangle_{\vec R,\vec P},
\end{equation}
where 
\begin{equation}
    \rho_\psi(\vec R,\vec P)=\int\frac{\ud^3q}{(2\pi)^3}\,e^{-i\vec q\cdot\vec R}\,\tilde\psi^*(\vec P+\tfrac{\vec q}{2})\tilde\psi(\vec P-\tfrac{\vec q}{2})
\end{equation}
with $\tilde\psi(\vec p)=\langle p|\psi\rangle/2\sqrt{p^0}$ is the Wigner distribution of the system~\cite{Wigner:1932eb}. It is then natural to interpret the amplitude
\begin{equation}
    \langle O\rangle_{\vec R,\vec P}=\int\frac{\ud^3\Delta}{(2\pi)^3}\,e^{i\vec\Delta\cdot\vec R}\,\frac{\langle p'|O|p\rangle}{2\sqrt{p^0p'^0}}
\end{equation}
as the expectation value of $O$ for a system localized in average around the position $\vec R$ with average momentum $\vec P=(\vec p'+\vec p)/2$. In the case of a local operator $O(x)$, one obtains a static distribution when there is no energy transfer to the system. Since the mass shell constraints imply $\Delta^0=\vec P\cdot\vec \Delta/P^0$, the elastic condition $\Delta^0=p'^0-p^0=0$ is automatically satisfied in the BF defined by $|\vec P|=0$. For a non-vanishing average momentum, one can choose the $z$-axis such that $\vec P=P_z\vec e_z$. The elastic condition is then satisfied provided one integrates over the longitudinal coordinate, leading to the two-dimensional (2D) EF distribution
\begin{equation}
    O_\text{EF}(\vec b_\perp;P_z)=\int\ud z\,\langle O\rangle_{\vec R,P_z\vec e_z}(\vec x)=\int\frac{\ud^2\Delta_\perp}{(2\pi)^2}\,e^{-i\vec\Delta_\perp\cdot\vec b_\perp}\,\frac{\langle p'|O(0)|p\rangle}{2P^0}\bigg|_{\Delta_z=|\vec P_\perp|=0}
\end{equation}
which depends on the relative transverse position $\vec b_\perp=\vec x_\perp-\vec R_\perp$ owing to translation symmetry.

The $P_z$-dependence of the EF distribution is essentially determined by the amplitude in momentum space. For a general tensor operator and spin-$j$ state, one expects that~\cite{Jacob:1959at,Durand:1962zza}
\begin{equation}
	\begin{aligned}\label{EFLorentzTrans-Spinj}
		&\langle p',s'|O^{\mu_1\cdots\mu_n}(0)|p,s\rangle = \\
  &\qquad\sum_{s_{B}',s_{B}} {D}^{*(j)}_{s_{B}'s'}(p_{B}',\Lambda){D}^{(j)}_{s_{B}s}(p_{B},\Lambda) \,\Lambda^{\mu_1}_{\phantom{\mu_1}\nu_1}\cdots \Lambda^{\mu_n}_{\phantom{\mu_n}\nu_n}\,\langle p_{B}',s_{B}'|O^{\nu_1\cdots\nu_n}(0)|p_{B},s_{B}\rangle,
	\end{aligned}
\end{equation}
where $\langle p_{B}',s_{B}'|O^{\nu_1\cdots\nu_n}(0)|p_{B},s_{B}\rangle$ is the BF matrix element, $\Lambda^{\mu_i}_{\phantom{\mu_i}\nu_i}$ is the Lorentz boost matrix from the BF to a generic Lorentz frame, and ${D}^{(j)}_{s_Bs}(p_B,\Lambda)$ is the corresponding Wigner spin rotation matrix. The latter contribution is the least familiar but is crucial for understanding the distortions induced by the boost~\cite{Lorce:2020onh,Lorce:2022jyi,Chen:2022smg,Chen:2023dxp}. In the following we will only consider a nucleon target, i.e. $j=\frac{1}{2}$.

%%%%%%%%%%%%%%%%%%%%%%%%%%%%%%%%%%%%%%%%%%%%%%%%%%%%%%%%%%%%%%%%%%%%%%%%%%%%%%%%%%
\section{Electromagnetic current}

According to the phase-space approach, the 3D BF distributions of electric charge and current are given by 
\begin{equation}
    J^\mu_B(\vec r)=\langle j^\mu\rangle_{\vec R,\vec 0}(\vec x)=\int\frac{\ud^3\Delta}{(2\pi)^3}\,e^{-i\vec\Delta\cdot\vec r}\,\frac{\langle p'_B,s'_B|j^\mu(0)|p_B,s_B\rangle}{2P^0_B}
\end{equation}
with $p^\mu_B=(P^0_B,-\vec\Delta/2)$ and $p'^\mu_B=(P^0_B,\vec\Delta/2)$ the initial and final momenta in the BF, and $\vec r=\vec x-\vec R$ the relative position. The difference with the conventional Sachs distributions~\cite{Ernst:1960zza,Sachs:1962zzc} lies in the normalization factor $2P^0_B$ instead of $2M$. Similarly, the corresponding 2D EF distributions are defined as
\begin{equation}
    J^\mu_\text{EF}(\vec b_\perp;P_z)=\int\ud z\,\langle j^\mu\rangle_{\vec R,P_z\vec e_z}(\vec x)=\int\frac{\ud^2\Delta_\perp}{(2\pi)^2}\,e^{-i\vec\Delta_\perp\cdot\vec b_\perp}\,\frac{\langle p',s'|j^\mu(0)|p,s\rangle}{2P^0}\bigg|_{\Delta_z=|\vec P_\perp|=0}.
\end{equation}
In the limit $P_z\to 0$, one obtains the projection of the BF distributions onto the transverse 2D plane
\begin{equation}
    J^\mu_\text{EF}(\vec b_\perp;0)=\int\ud z\,J^\mu_B(\vec r).
\end{equation}
Moreover, the total electric charge
\begin{equation}
    \mathcal Q=\int\ud^2b_\perp\, J^0_\text{EF}(\vec b_\perp;P_z)\big|_{s'=s}=\frac{\langle p,s|j^0(0)|p,s\rangle}{2p^0}
\end{equation}
does not depend on $P_z$, showing that the amplitudes in momentum space are correctly normalized.

Using the traditional parametrization in terms of the Dirac and Pauli FFs
\begin{equation}\label{genparam}
	\langle p',s'|j^\mu(0)|p,s\rangle=e\,\overline u(p',s')\left[\gamma^\mu\,F_1(Q^2)+\frac{i\sigma^{\mu\nu}\Delta_\nu}{2M}\,F_2(Q^2)\right]u(p,s),
\end{equation}
with $Q^2=-\Delta^2$, one finds in the BF the same spin structure as in the non-relativistic case~\cite{Yennie:1957skg,Ernst:1960zza,Sachs:1962zzc}
\begin{equation}\label{BFampl}
	\begin{aligned}
		\langle p'_B,s'_B|j^0(0)|p_B,s_B\rangle&=e\,2M\,\delta_{s'_Bs_B}\,G_E(Q^2),\\
		\langle p'_B,s'_B|\vec j(0)|p_B,s_B\rangle&=e\,(\vec\sigma_{s'_Bs_B}\times i\vec\Delta)\,G_M(Q^2),
	\end{aligned}
\end{equation}
where $G_E=F_1-\tau F_2$ with $\tau=Q^2/(4M^2)$ and $G_M=F_1+F_2$ are the electric and magnetic Sachs FFs, and $\vec\sigma$ are the three Pauli matrices. 

The spin structure in a generic EF is more complicated~\cite{Lorce:2020onh,Kim:2021kum,Chen:2022smg} but can be massaged in the following form
\begin{equation}\label{spinhalfexplicitLT}
    \begin{aligned}
        \langle p',s'|j^0(0)|p,s\rangle\big|_\text{EF}&=e\,2M\,\gamma\,\Bigg[\left(\delta_{s's}\,\cos\theta+\frac{(\vec\sigma_{s's}\times i\vec \Delta_\perp)_z}{|\vec\Delta_\perp|}\,\sin\theta\right)G_E(Q^2)\\
        &\qquad\qquad\quad +\beta\left(-\delta_{s's}\,\sin\theta+\frac{(\vec\sigma_{s's}\times i\vec \Delta_\perp)_z}{|\vec\Delta_\perp|}\,\cos\theta\right)\sqrt{\tau}\,G_M(Q^2)\Bigg],\\
         \langle p',s'|j^3(0)|p,s\rangle\big|_\text{EF}&=e\,2M\,\gamma\,\Bigg[\beta\left(\delta_{s's}\,\cos\theta+\frac{(\vec\sigma_{s's}\times i\vec \Delta_\perp)_z}{|\vec\Delta_\perp|}\,\sin\theta\right)G_E(Q^2)\\
        &\qquad\qquad\quad +\left(-\delta_{s's}\,\sin\theta+\frac{(\vec\sigma_{s's}\times i\vec \Delta_\perp)_z}{|\vec\Delta_\perp|}\,\cos\theta\right)\sqrt{\tau}\,G_M(Q^2)\Bigg],\\
        \langle p',s'|\vec j_\perp(0)|p,s\rangle\big|_\text{EF}&=e\,(\sigma_z)_{s's}\,(\vec e_z\times i\vec \Delta_\perp)\,G_M(Q^2),
    \end{aligned}
\end{equation}
in agreement with the general expectation~\eqref{EFLorentzTrans-Spinj}. The Lorentz boost parameters are
\begin{equation}\label{boostparam}
    \gamma=\frac{P^0}{\sqrt{P^2}}=\frac{P^0}{P^0_B}=\frac{P^0}{M\sqrt{1+\tau}},\qquad \beta=\frac{P_z}{P^0},
\end{equation}
and the Wigner rotation angle $\theta$ satisfies
\begin{equation}\label{WignerAngleSinCos}
    \cos\theta=\frac{P^0+M(1+\tau)}{(P^0+M)\sqrt{1+\tau}},\qquad \sin\theta=-\frac{\sqrt{\tau}P_z}{(P^0+M)\sqrt{1+\tau}}.
\end{equation}
In the BF limit $P_z\to 0$, one recovers~\eqref{BFampl} restricted to $\Delta_z=0$. In the IMF limit $P_z\to\infty$, one obtains using the LF coordinates $a^\pm=(a^0\pm a^3)/\sqrt{2}$
\begin{equation}\label{LFspinhalfexplicitPM}
    \begin{aligned}
        \langle p',s'|j^+(0)|p,s\rangle\big|_\text{IMF}&=e\,2P^+\,\bigg[\delta_{s's}\,F_1(Q^2)+\frac{(\vec\sigma_{s's}\times i\vec\Delta)_z}{2M}\,F_2(Q^2)\bigg],\\
         \langle p',s'|j^-(0)|p,s\rangle\big|_\text{IMF}&=e\,2P^-\,\bigg[\delta_{s's}\,G_1(Q^2)+\frac{(\vec\sigma_{s's}\times i\vec\Delta)_z}{2M}\,G_2(Q^2)\bigg],\\
        \langle p',s'|\vec j_\perp(0)|p,s\rangle\big|_\text{IMF}&=e\,(\sigma_z)_{s's}\,(\vec e_z\times i\vec \Delta_\perp)\,G_M(Q^2),
    \end{aligned}
\end{equation}
where $G_1=(G_E-\tau G_M)/(1+\tau)$ and $G_2=-(G_E+G_M)/(1+\tau)$. These amplitudes agree with those defining the 2D distributions within the LF formalism~\cite{Burkardt:2002hr,Miller:2007uy,Carlson:2007xd}. In particular, the canonical polarization in the IMF coincides with the LF helicity.

\begin{figure}[t]
\includegraphics[width=\hsize]{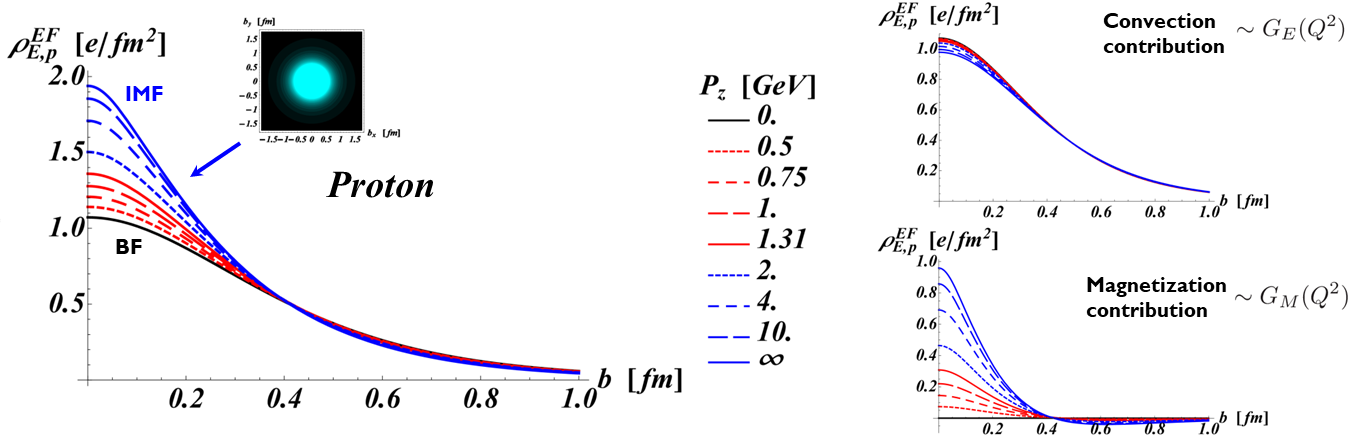}
\includegraphics[width=\hsize]{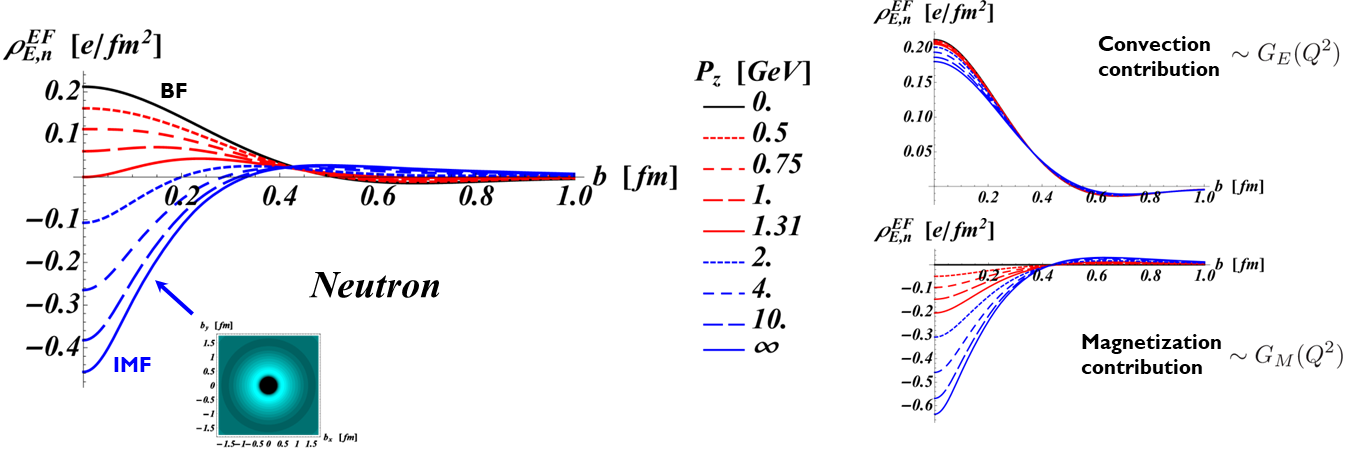}
\caption{Unpolarized EF charge distributions $\rho^{EF}_E\equiv J^0_\text{EF}(\vec b_\perp;P_z)$ in the proton and the neutron for different values of the target average momentum. The small panels on the right show the electric (or convective) and magnetic (or magnetization) contributions. Figure adapted from~\cite{Lorce:2020onh}.}\label{fig:rhoE}
\end{figure}

Clearly, the effect of a boost is not limited to a mere mixing of the $j^0$ and $j^3$ components of the electromagnetic current, but induces also a distortion of the spin structure. As a result, both the electric and magnetic terms contribute to the unpolarized charge distribution as soon as $\vec P\neq \vec 0$. The fact that the induced magnetic contribution in the neutron is large and opposite to the electric contribution explains in particular why the center of the LF charge distribution of the neutron appears to be negative, see Fig.~\ref{fig:rhoE}. Because of these kinematical distortions one cannot interpret LF distributions as intrinsic densities, although they are probabilistic. 

%%%%%%%%%%%%%%%%%%%%%%%%%%%%%%%%%%%%%%%%%%%%%%%%%%%%%%%%%%%%%%%%%%%%%%%%%%%%%%%%%%
\section{Energy-momentum tensor}

The same formalism can be applied to the energy-momentum tensor (EMT) to discuss nucleon mechanical properties. The concept of EMT distributions in the BF was introduced in~\cite{Polyakov:2002yz} and generalized to both the EF and the LF in~\cite{Lorce:2018egm}.

The spin-$\frac{1}{2}$ matrix elements of the quark EMT can conveniently be parametrized as follows
\begin{equation}
\begin{aligned}
     \langle p',s'|T^{\mu\nu}_q(0)|p,s\rangle&=\overline u(p',s')\left[\frac{P^\mu P^\nu}{M}\,A_q(Q^2)+\frac{\Delta^\mu\Delta^\nu-g^{\mu\nu}\Delta^2}{4M}\,D_q(Q^2)+Mg^{\mu\nu}\bar C_q(Q^2)\right.\\
     &\qquad\qquad\quad+\left.\frac{P^{\{\mu}i\sigma^{\nu\}\lambda}\Delta_\lambda}{M}\,J_q(Q^2)-\frac{P^{[\mu}i\sigma^{\nu]\lambda}\Delta_\lambda}{M}\,S_q(Q^2)\right]u(p,s),
\end{aligned}
\end{equation}
and similarly for the gluon EMT. The first four terms describe the symmetric part of the EMT $T^{\{\mu\nu\}}=(T^{\mu\nu}+T^{\nu\mu})/2$ while the last term accounts for the antisymmetric part $T^{[\mu\nu]}=(T^{\mu\nu}-T^{\nu\mu})/2$. Poincar\'e symmetry implies some constraints on the so-called gravitational FFs
\begin{equation}
        A_q(0)+A_g(0)=1,\qquad
        J_q(0)+J_g(0)=\tfrac{1}{2},\qquad \bar C_q(Q^2)+\bar C_g(Q^2)=0,
\end{equation}
which can be understood in terms of linear momentum conservation, angular momentum conservation, and mechanical equilibrium, respectively.

\begin{figure}[t]
\centering\includegraphics[width=.55\hsize]{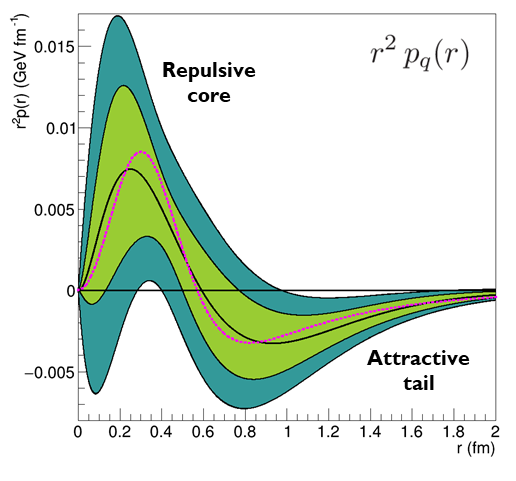}
\caption{Radial distribution of the quark pressure $r^2 p_q(r)$ in the proton based on JLab data~\cite{Burkert:2018bqq}. Note that the contribution from $\bar C_q(Q^2)$ is currently not known and was set to zero. Figure adapted from~\cite{Burkert:2023wzr}.}\label{fig:p}
\end{figure}

To keep the presentation simple, the following discussion is restricted to the BF. One finds
\begin{equation}\label{BFEMTampl}
	\begin{aligned}
		\langle p'_B,s'_B|T^{00}_q(0)|p_B,s_B\rangle&=2MP^0_B\,\delta_{s'_Bs_B}\left\{A_q(Q^2)+\bar C_q(Q^2)+\tau[D_q(Q^2)-B_q(Q^2)]\right\},\\
		\langle p'_B,s'_B|T^{\{0k\}}_q(0)|p_B,s_B\rangle&=2P^0_B\,(\vec\sigma_{s'_Bs_B}\times i\vec\Delta)^k\,J_q(Q^2),\\
  \langle p'_B,s'_B|T^{[0k]}_q(0)|p_B,s_B\rangle&=-2P^0_B\,(\vec\sigma_{s'_Bs_B}\times i\vec\Delta)^k\,S_q(Q^2),\\
  \langle p'_B,s'_B|T^{ij}_q(0)|p_B,s_B\rangle&=2MP^0_B\,\delta_{s'_Bs_B}\left\{\frac{\Delta^i\Delta^j}{4M^2}\,D_q(Q^2)-\delta^{ij}[\bar C_q(Q^2)+\tau D_q(Q^2)]\right\},
	\end{aligned}
\end{equation}
where $B_q=2J_q-A_q$. These matrix elements are key to the physics program of the future Electron-Ion Collider~\cite{AbdulKhalek:2021gbh} for they encode a lot of information about a large number of fundamental properties of the nucleons. Here are a few examples:
\begin{itemize}
    \item The normalized amplitude $\langle p'_B,s'_B|T^{00}_q(0)|p_B,s_B\rangle/(2P^0_B)$ describes after Fourier transform the spatial distribution of the quark energy $T^{00}_{q,B}(\vec r)$, and reduces in the forward limit $\Delta\to 0$ to the quark contribution to the nucleon mass~\cite{Lorce:2017xzd,Lorce:2021xku}. 
    \item The spatial distribution of quark Belinfante angular momentum is defined as
\begin{equation}
    \mathcal J^i_{q,B}(\vec r)=\epsilon^{ijk}r^j T^{\{0k\}}_{q,B}(\vec r)
\end{equation}
and differs from the spatial distribution of quark kinetic angular momentum~\cite{Leader:2013jra,Lorce:2017wkb}
\begin{equation}
    J^i_{q,B}(\vec r)=L^i_{q,B}(\vec r)+S^i_{q,B}(\vec r),
\end{equation}
where orbital and spin contributions read
\begin{equation}
    \begin{aligned}
        L^i_{q,B}(\vec r)=\epsilon^{ijk}r^j T^{0k}_{q,B}(\vec r),\qquad
        S^i_{q,B}(\vec r)=\frac{1}{2}\langle\overline\psi\gamma^i\gamma_5\psi\rangle_{\vec 0,\vec 0}(\vec r). 
    \end{aligned}
\end{equation}

\item The stress tensor is of particular interest~\cite{Polyakov:2002yz,Polyakov:2018zvc,Lorce:2018egm}
\begin{equation}
    T^{ij}_{q,B}(\vec r)=\delta^{ij}\,p_q(r)+\left(\frac{r^i r^j}{r^2}-\frac{1}{3}\,\delta^{ij}\right)s_q(r).
\end{equation}
Once summed over quark and gluon contributions, isotropic pressure $p(r)$ and pressure anisotropy (or shear forces) $s(r)$ are related via the EMT conservation
\begin{equation}
    \nabla^iT^{ij}_B(\vec r)=0\qquad\Rightarrow\qquad \frac{\ud}{\ud r}\bigg(p+\frac{2}{3}\,s\bigg)+\frac{2}{r}\,s=0.
\end{equation}
As a corollary of the virial theorem, the von Laue condition $\int\ud^3r\,p(r)=0$ is satisfied to ensure mechanical equilibrium of the system. It has also been suggested that the quantity $D_q(0)+D_g(0)=M\int\ud^3r\,r^2\,p(r)$ should be negative for stability reasons~\cite{Perevalova:2016dln}. 
\end{itemize}
The first experimental extraction of the pressure distribution inside the proton from deeply virtual Compton scattering appeared in~\cite{Burkert:2018bqq}, see Fig.~\ref{fig:p}. More conservative analyses were later reported in~\cite{Kumericki:2019ddg,Dutrieux:2021nlz}. Gluon gravitational FFs have also been extracted from the photoproduction of $J/\psi$ in the threshold region~\cite{Duran:2022xag}. On the Lattice QCD side, recent estimates of the gravitational FFs can be found e.g.~in~\cite{Alexandrou:2020sml,Pefkou:2021fni,Wang:2021vqy}. For more details and references, see the review~\cite{Burkert:2023wzr}.

%%%%%%%%%%%%%%%%%%%%%%%%%%%%%%%%%%%%%%%%%%%%%%%%%%%%%%%%%%%%%%%%%%%%%%%%%%%%%%%%%%
\section{Conclusions}

Electromagnetic and gravitational form factors are Lorentz-invariant functions encoding the information about the spatial structure of a system. In a relativistic regime, recoil effects cannot in general be neglected. As a result, spatial distributions are usually frame-dependent. The phase-space formalism is therefore the natural language to formulate the relativistic notion of spatial distribution, albeit in a quasi-probabilistic picture. This approach interpolates between the Breit frame ($P_z=0$) and light-front ($P_z\to\infty$) representations, and clearly shows that the spin of the target plays a key role when it comes to understanding how spatial distributions are affected by a Lorentz boost.

\end{document}